\documentclass[aps,prl,twocolumn,superscriptaddress,amsmath,amssymb]{revtex4}
\usepackage{float}
\usepackage{makeidx}
\usepackage{graphicx,amsmath}
\usepackage{CJK}
\usepackage{colordvi}
\usepackage{color}
\usepackage{appendix}

\begin{document}
\begin{CJK*}{GB}{gbsn}

\title{Enhanced sensitivity of low-frequency signal by using broad squeezed light and bichromatic local oscillator}

\author{Wei Li}

\affiliation{State Key Laboratory of Quantum Optics and Quantum
Optics Devices, Institute of Opto-Electronics, Shanxi University,
Taiyuan 030006, P.R.China }

\affiliation{Collaborative Innovation Center of Extreme Optics,
Shanxi University, Taiyuan 030006, P.R.China}

\author{Yuanbin Jin}

\affiliation{State Key Laboratory of Quantum Optics and Quantum
Optics Devices, Institute of Opto-Electronics, Shanxi University,
Taiyuan 030006, P.R.China }

\affiliation{Collaborative Innovation Center of Extreme Optics,
Shanxi University, Taiyuan 030006, P.R.China}

\author{Xudong Yu}
\affiliation{State Key Laboratory of Quantum Optics and Quantum
Optics Devices, Institute of Opto-Electronics, Shanxi University,
Taiyuan 030006, P.R.China }

\affiliation{Collaborative Innovation Center of Extreme Optics,
Shanxi University, Taiyuan 030006, P.R.China}

\author{Jing Zhang,$^{\dagger}$}

\affiliation{State Key Laboratory of Quantum Optics and Quantum
Optics Devices, Institute of Opto-Electronics, Shanxi University,
Taiyuan 030006, P.R.China }

\affiliation{Synergetic Innovation Center of Quantum Information and
Quantum Physics, University of Science and Technology of China,
Hefei, Anhui 230026, P. R. China}

\begin{abstract}
We experimentally study a protocol of using the broadband
high-frequency squeezed vacuum to detect the low-frequency signal.
In this scheme, the lower sideband field of the squeezed light
carries the low-frequency modulation signal and the two strong
coherent light fields are applied as the bichromatic local
oscillator in the homodyne detection to measure the quantum
entanglement of the upper and lower sideband for the broadband
squeezed light. The power of one of local oscillators for detecting
the upper sideband can be adjusted to optimize the conditional
variance in the low frequency regime by subtracting the photocurrent
of the upper sideband field of the squeezed light from that of the
lower sideband field. By means of the quantum correlation of the
upper and lower sideband for the broadband squeezed light, the
low-frequency signal beyond the standard quantum limit is measured.
This scheme is appropriate for enhancing sensitivity of
low-frequency signal by the aid of the broad squeezed light, such as
gravitational waves detection, and does not need to directly produce
the low frequency squeezing in optical parametric process.

\end{abstract}
\maketitle
\end{CJK*}

The first detection of gravitational waves (GW) emitted from the
merger of two black holes by the Laser Interferometer
Gravitational-Wave Observatory (LIGO) sets the course of a new era
of astrophysics. GW detection is now opening an exciting new
observational frontier in astronomy and cosmology
\cite{Abbott2016-1,Abbott2016-2,Abbott2016-3}. The further
improvement of GW detector sensitivity is expected to extend the
detection range and the event rate of binary black holes
coalescence, and may lead to detections of more exotic sources.

In Advanced LIGO, vacuum fluctuations entering from the dark port of
the interferometer \cite{Caves1981} can make the quadrature phase of
the output carrier field at the dark port noisy, while which
contains GW signal. However as the squeezed vacuum state is fed into
the dark port of the interferometer, the sensitivity can be improved
beyond the standard quantum limit (SQL)
\cite{Xiao1987,Grangier1987}. The use of squeezed states to enhance
the sensitivity began with initial proof-of-principle experiments
and recently have been demonstrated in GEO 600
\cite{LIGO-squeeze2011} and LIGO \cite{LIGO-squeeze2013}. Since
terrestrial GW signal locates in 10 Hz to 10 kHz band
\cite{Kimble2001, Chelkowski2005}, the squeezing in the audio band
is required, which is a great technical challenge. Until now, there
is a very wide research demonstrating the squeezing at the lower
frequency band \cite{Fabre,Lett,Mikhailov} and applications in
quantum metrology
\cite{Lawrie,Lawrie2,Pooser,Andersen,Marino2016,Lawrie3,Andersen2,Lehnert,Lett2,Pooser2,Fabre2}.

Broadband squeezing has been demonstrated at megahertz frequencies,
where technical noise sources of the laser light are not present. At
these frequencies, the laser operates at or near the shot-noise
limit. Due to the strong quantum correlation between the lower and
upper sideband of the squeezed light field, the single broadband
squeezed light can be split into N pairs of upper and lower sideband
fields with spatial separation to produce N independent
Einstein-Podolsky-Rosen (EPR) entangled fields \cite{jzhang}. This
scheme was demonstrated experimentally by using a pair of
frequency-shifted local oscillators to measure this EPR entanglement
\cite{Huntington,Hage}. Recently, a theoretical protocol is proposed
to improve LIGO's sensitivity beyond the SQL via EPR entanglement of
the broad squeezing field and the dual use of the interferometer as
both the GW detector and the filter, eliminating the need for
external narrow filter cavities \cite{Ma2016}. In this paper, we
employ a broadband high-frequency squeezed vacuum to detect
low-frequency signal beyond the standard quantum limit. The
broadband squeezed vacuum consists a pair of EPR entangled beams:
the signal beam (lower sideband field) around the carrier frequency
$\omega_{0}$, and the idler beam (upper sideband field) around
$\omega_{0}+\Lambda$. The lower sideband field around the carrier
frequency $\omega_{0}$ will carry the low-frequency modulation
signal around the carrier frequency $\omega_{0}$, however, the upper
sideband field around $\omega_{0}+\Lambda$ feel nothing. The output
lower and upper sideband fields may be separated in space by a mode
cleaner cavity and measured by homodyne detection with two local
oscillators at frequency $\omega_{0}$ and $\omega_{0}+\Lambda$
respectively. The conditional squeezing of the output signal beam
can be obtained by subtracting the photocurrent of the idler beam
from that of the signal beam. Here, the lower and upper sideband
fields of the broad squeezed light may be separated in space by a
mode cleaner cavity before or after carrying the low-frequency
modulation signal. Thus this scheme also can be considered as: 1)
First, the broadband squeezed vacuum is separated into the lower and
upper sideband fields in space by a mode cleaner cavity. 2) The
lower sideband field around the carrier frequency $\omega_{0}$ is
sent into sensitive device (such as the interferometer), therefore,
the low-frequency signal around the carrier frequency $\omega_{0}$
is added in the lower sideband field by the sensitive device. 3) The
lower and upper sideband fields are measured by homodyne detection
with two local oscillators at frequency $\omega_{0}$ and
$\omega_{0}+\Lambda$ respectively. The conditional variance of the
lower sideband beam can be obtained by subtracting the photocurrent
of the upper sideband beam from that of the lower sideband beam.
Thus this scheme can avoid producing the low frequency squeezing to
improve the sensitivity of the interferometer.

In this paper, we remove the mode cleaner cavity to separate the
signal and idler beams in space and utilize a bichromatic  local
oscillator (BLO) to directly detect the signal and idler beams of a
broad high-frequency squeezed vacuum after carrying the
low-frequency signal by combining with a phase-modulated coherent
light at around $\omega_{0}$ on a beam splitter. This scheme can
avoid optical losses introduced by the mode cleaner cavity.
Moreover by optimally adjusting the power of one of local oscillator
for detecting the idler field, we can obtain the minimum conditional
variance of the signal beam and improve the signal noise ratio. The
theoretical scheme based on BLO to detect the squeezed state was
proposed \cite{Marino} and the phase-sensitive detection with a BLO
or a double-sideband signal field were studied experimentally
\cite{Fan,LiW,appel}. And the measurement of a broad squeezed vacuum
state by means of a BLO was demonstrated experimentally
\cite{LiW2015,li2017}. Recently, by making use of the multi-frequency
homodyne detection, the experiments of cross-frequency entanglements
generated in periodically pumped OPOs have been reported
\cite{Pfister, Treps}.

The schematic diagram of the detection is shown in Fig. 1(a). A BLO
with two local oscillators at frequency $\omega_{0}$ and
$\omega_{0}+\Lambda$ is mixed with the detected light field at a
50/50 beam splitter. The power of one of local oscillator (upper
local oscillator) at $\omega_{0}+\Lambda$ can be adjusted with the
factor g. The relative phase $\theta$ of the local oscillator and
the detected field can be controlled by the reflective mirror
mounted on a PZT (piezoelectric transducer). The annihilation
operators of the BLO and the detected signal field can be written as
$\hat{a}(t)=\hat{a}_{-}(t)\exp[-i\omega_{0}t]+\hat{a}_{+}(t)\exp[-i(\omega_{0}+\Lambda)t]$
and $\hat{b}(t)=\hat{b}_{0}(t)\exp{(-i\omega_{0}{t})}$, where
$\hat{a}_{+(-)}(t)$ and $\hat{b}_{0}(t)$ are the slow varying
operators of the fields. The normalized difference of the
photocurrents of the two detectors at the 50/50 beam splitter is
\begin{eqnarray}
\hat{i}(t)=\frac{1}{a}[\langle\hat{a}^{\dag}(t)\rangle\hat{b}(t)e^{-i\theta}+\langle\hat{a}(t)\rangle\hat{b}^{\dag}(t)
e^{i\theta}],
\end{eqnarray}
where the fields satisfy $\langle\hat{a}_{+} \rangle=ga$, and
$\langle\hat{a}_{-}\rangle=$ $a\gg\langle\hat{b}_{0}\rangle\sim0$.
Therefore the BLO is a pair of the strong coherent states. And the
detected field is the vacuum state or the squeezed vacuum state
carrying the low-frequency signal around frequency $\omega_{0}$.

The difference of the photocurrents analyzed at the radio frequency
$\Omega$ is expressed as
\begin{eqnarray}
\hat{i}(\Omega)&=&\hat{Q}_{-}(\Omega,\theta)+g\hat{Q}_{+}(\Omega,\theta).
\end{eqnarray}
Here, we express the quadrature component of the signal field around
the central frequency $\omega_{0}$, which easily compare with the
measurement with a single local oscillator ($g=0$) at $\omega_{0}$.
Therefore, the quadrature component of the detected field can be
defined as
$\hat{Q}_{-}(\Omega,\theta)=\hat{b}(\omega_{0}-\Omega)e^{-i\theta}+\hat{b}^{\dag}(\omega_{0}+\Omega)e^{i\theta}$,
and
$\hat{Q}_{+}(\Omega,\theta)=\hat{b}(\omega_{0}+\Lambda-\Omega)e^{-i\theta}+\hat{b}^{\dag}(\omega_{0}+\Lambda+\Omega)e^{i\theta}$.
The quadrature amplitude ($\theta=0$) can be
$\hat{X}_{-}(\Omega)=\hat{b}(\omega_{0}-\Omega)+\hat{b}^{\dag}(\omega_{0}+\Omega)$
and the quadrature phase ($\theta=\pi/2$)
$\hat{Y}_{-}(\Omega)=-i[\hat{b}(\omega_{0}-\Omega)-\hat{b}^{\dag}(\omega_{0}+\Omega)]$.
The arbitrary quadrature component of the detected field can be
measured by scanning the relative phase of $\theta$. So when
$\theta=0$, the difference of the photocurrents will give the
information of the quadrature amplitude of the detected field
$\hat{X}_{B}(\Omega)=\hat{X}_{-}(\Omega)+g\hat{X}_{+}(\Omega)$, and
when $\theta=\pi/2$, the quadrature phase
$\hat{Y}_{B}(\Omega)=\hat{Y}_{-}(\Omega)+g\hat{Y}_{+}(\Omega)$.

Since a single broadband squeezed light can be split into a pair of
upper and lower sideband fields as EPR entangled fields, the minimum
conditional variance of the output lower sideband (signal) beam can
be obtained with the help of the upper sideband (idler) beam.
Considering the simple optical parametric oscillator (OPO) process,
the nonlinear medium is pumped with the second-harmonic wave of
$\omega_{p}=2\omega_{0}+\Lambda$. The annihilation operators of the
output lower and upper sideband fields of OPO can be written as
\begin{eqnarray}
\hat{b}_{-}^{s}&=&\hat{b}_{-}^{0}\cosh{r}+\hat{b}_{+}^{\dagger 0}e^{i\theta_{p}}\sinh{r},\nonumber\\
\hat{b}_{+}^{s}&=&\hat{b}_{+}^{0}\cosh{r}+\hat{b}_{-}^{\dagger
0}e^{i\theta_{p}}\sinh{r},
\end{eqnarray}
where $r$ is squeezing factor, $\theta_{p}$ is the phase of the pump
field, $\hat{b}_{-}^{0}$ and $\hat{b}_{+}^{0}$ are the annihilation
operators of the input lower and upper sideband vacuum fields of OPO
with
$\langle\delta^{2}X_{-}^{0}(\Omega)\rangle=\langle\delta^{2}X_{+}^{0}(\Omega)\rangle=\langle\delta^{2}Y_{-}^{0}(\Omega)\rangle=\langle\delta^{2}Y_{+}^{0}(\Omega)\rangle=1$.
If $\theta_{p}=0$, the phase amplitudes of the output lower and
upper sideband fields of OPO can be given by
\begin{eqnarray}
\hat{X}_{-}^{s}(\Omega)&=&\hat{X}_{-}^{0}(\Omega)\cosh{r}+\hat{X}_{+}^{0}(\Omega)\sinh{r},\nonumber\\
\hat{Y}_{-}^{s}(\Omega)&=&\hat{Y}_{-}^{0}(\Omega)\cosh{r}-\hat{Y}_{+}^{0}(\Omega)\sinh{r},\nonumber\\
\hat{X}_{+}^{s}(\Omega)&=&\hat{X}_{+}^{0}(\Omega)\cosh{r}+\hat{X}_{-}^{0}(\Omega)\sinh{r},\nonumber\\
\hat{Y}_{+}^{s}(\Omega)&=&\hat{Y}_{+}^{0}(\Omega)\cosh{r}-\hat{Y}_{-}^{0}(\Omega)\sinh{r},
\end{eqnarray}
then, the difference and sum of amplitude phase quadratures of the
output lower and upper sideband fields of OPO are obtained
\begin{eqnarray}
\hat{X}_{-}^{s}(\Omega)-\hat{X}_{+}^{s}(\Omega)&=&(\hat{X}_{-}^{0}(\Omega)-\hat{X}_{+}^{0}(\Omega))e^{-r},\nonumber\\
\hat{Y}_{-}^{s}(\Omega)+\hat{Y}_{+}^{s}(\Omega)&=&(\hat{Y}_{-}^{0}(\Omega)+\hat{Y}_{+}^{0}(\Omega))e^{-r},\nonumber\\
\hat{X}_{-}^{s}(\Omega)+\hat{X}_{+}^{s}(\Omega)&=&(\hat{X}_{-}^{0}(\Omega)+\hat{X}_{+}^{0}(\Omega))e^{+r},\nonumber\\
\hat{Y}_{-}^{s}(\Omega)-\hat{Y}_{+}^{s}(\Omega)&=&(\hat{Y}_{-}^{0}(\Omega)-\hat{Y}_{+}^{0}(\Omega))e^{+r}.
\end{eqnarray}
The variances of the output lower and upper sideband fields of OPO
are expressed by
\begin{eqnarray}
\langle\delta^{2}X_{-}^{s}(\Omega)\rangle&=&\langle\delta^{2}X_{+}^{s}(\Omega)\rangle=\langle\delta^{2}Y_{-}^{s}(\Omega)\rangle=\langle\delta^{2}Y_{+}^{s}(\Omega)\rangle \nonumber\\
&=&\frac{e^{-2r}+e^{2r}}{2},
\end{eqnarray}
and the correlated variances are given by
\begin{eqnarray}
\langle\delta^{2}(\hat{X}_{-}^{s}(\Omega)-\hat{X}_{+}^{s})(\Omega)\rangle&=&\langle\delta^{2}(\hat{Y}_{-}^{s}(\Omega)+\hat{Y}_{+}^{s}(\Omega))\rangle=2e^{-2r},\nonumber\\
\langle\delta^{2}(\hat{X}_{-}^{s}(\Omega)+\hat{X}_{+}^{s})(\Omega)\rangle&=&\langle\delta^{2}(\hat{Y}_{-}^{s}(\Omega)-\hat{Y}_{+}^{s}(\Omega))\rangle=2e^{+2r}.
\end{eqnarray}

The variance of the conditional quadrature phase
$\hat{Y}_{B}(\Omega)$ detected by BLO with the factor g is expressed
by
\begin{eqnarray}
\langle\delta^{2}\hat{Y}_{B}(\Omega)\rangle&=&\langle\delta^{2}(\hat{Y}_{-}^{s}+g\hat{Y}_{+}^{s})\rangle\nonumber\\
&=&\frac{e^{2r}}{2}(1-g)^{2}+\frac{e^{-2r}}{2}(1+g)^{2}.
\end{eqnarray}
In parallel, the conditional quadrature amplitude
$\hat{X}_{B}(\Omega)$ is given by
\begin{eqnarray}
\langle\delta^{2}\hat{X}_{B}(\Omega)\rangle&=&\langle\delta^{2}(\hat{X}_{-}^{s}+g\hat{X}_{+}^{s})\rangle\nonumber\\
&=&\frac{e^{-2r}}{2}(1-g)^{2}+\frac{e^{2r}}{2}(1+g)^{2}.
\end{eqnarray} When we choose the optimized value of
$g_{opt}=(e^{2r}-e^{-2r})/(e^{2r}+e^{-2r})$, the minimum conditional
variance of the output lower sideband field is obtained
\begin{eqnarray}
\langle\delta^{2}\hat{Y}_{B}^{opt}(\Omega)\rangle&=&\frac{2}{e^{2r}+e^{-2r}}.
\end{eqnarray}
Here, the method of obtaining the minimum conditional variance by
the optimized factor is same as the previous works
\cite{Reid,Lett3}. Thus, for the vacuum field injection, $g=0$
(without the upper local oscillator $\omega_{0}+\Lambda$) and
$\langle\delta^{2}\hat{Y}_{B}(\Omega)\rangle=1$. When the broadband
squeezed light with 3 dB is injected, the minimum conditional
variance is -0.97 dB with $g_{opt}=0.6$.

\begin{figure}
\centerline{
\includegraphics[width=8cm]{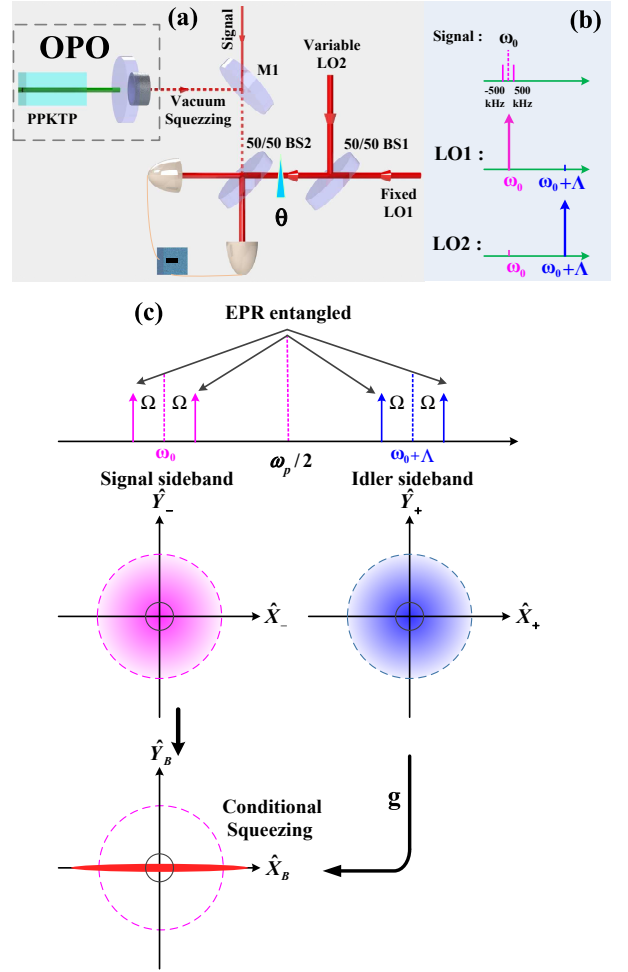}} \vspace{0.1in}
%\setlength{\columnwidth}{3.2in}
%\centerline{
\caption{\textbf{The experimental setup and schematic diagram of
detecting low-frequency signal beyond the standard quantum limit by
a broadband squeezing and BLO.} \textbf{a}, The experimental setup.
OPO, optical parametric oscillator; BS, 50/50 beam splitter; LO,
local oscillator; M1, 98/2 beam splitter. \textbf{b}, Spectra of the
weak low-frequency signal and the BLO. \textbf{c}, Spectral
decomposition of EPR-entanglement for the broad squeezed light
(upper panel) and prepare the conditional squeezing of the signal
beam by BLO detection (lower panel). \label{Fig1} }
%}
\end{figure}

The experimental setup and schematic diagram are shown in Fig. 1. A
diode-pumped intra-cavity frequency-doubled single-frequency laser
provides the fundamental light of 200 mW at 1064 nm and the
second-harmonic light of 450 mW at 532 nm simultaneously. The
second-harmonic light with the frequency of
$\omega_{p}=2\omega_{0}+\Lambda$ is used to pump an OPO to generate
the broad squeezed vacuum field. The OPO cavity is resonant for both
the pump light at 532 nm and the fundamental light at 1064 nm. The
OPO cavity with 38 mm long contains a type-I PPKTP crystal (1
mm$\times$2 mm$\times$10 mm), the front facet of which is highly
reflective for 1064 nm and has a power transmittance of 5$\%$ for
532 nm, and an outcoupling mirror that is highly reflective at 532
nm and has an intensity transmittance of 12.5$\%$ at 1064 nm. The
bandwidth of the OPO is about 70 MHz. The OPO cavity is locked by
PDH (Pound-Driver-Hall) technology and the error signal is extracted
by detecting the reflected pump light of the OPO. The output broad
squeezed field carries the weak low-frequency signal ($\pm 500 kHz$)
around the carrier frequency $\omega_{0}$ at M1 (98/2 beam
splitter).

The fundamental output field ($\omega_{p}/2$) passes through the
acousto-optical frequency-shifted system and then is split into two
beams with the frequency of $\omega_{0}=\omega_{p}/2-5 MHz$ and
$\omega_{0}+\Lambda=\omega_{p}/2+5 MHz$ (here, $\Lambda=10 MHz$). In
the frequency-shifted system, AOM1 shifts the laser frequency by the
first-order diffraction with a mount of +110 MHz. Then the
frequency-shifted laser is split into two parts, which are
translated back by AOM2 and AOM3 with the mount of -105 MHz and -115
MHz respectively. The two frequency-shifted laser beams at
$\omega_{p}/2\pm5 MHz$ as local oscillator (LO) 1 and 2 are combined
on 50/50 BS with the same polarization to generate the BLO. Here,
the power of LO1 is fixed and that of LO2 can be varied. A small
portion from the frequency-shifted laser beams at $\omega_{p}/2-5
MHz$ is used to generate the weak low-frequency signal by a phase
modulator. In order to lock the relative frequency and phase of the
two LOs, the signal generators of the acousto-optical
frequency-shifted system are locked by the clock synchronization
technology \cite{LiW,LiW2015}. The squeezed light with the weak
low-frequency signal is mixed with the BLO on the 50/50 BS. Finally,
the two output fields of the BS are detected by two balanced
detectors.

\begin{figure}
\centerline{
\includegraphics[width=8cm]{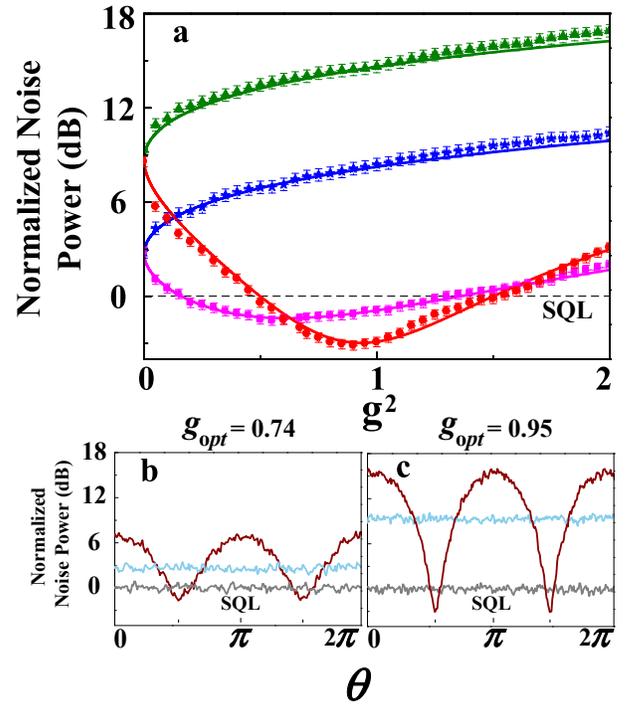}} \vspace{0.1in}
%\setlength{\columnwidth}{3.2in}
%\centerline{
\caption{\textbf{The noise variance of the conditional quadrature
phase amplitude.} \textbf{a}, The noise variance of the conditional
quadrature phase amplitude as the function of the factor $g^{2}$.
The conditional quadrature phase (magenta squares) and amplitude (blue
stars) are for the initial squeezing 3.9 dB (antisqueezing 5.24 dB)
and the extra noise $N_{e}=1.75$. The conditional quadrature phase
(red circles) and amplitude (green triangles) are for the initial
squeezing 5.9 dB (antisqueezing 11.6 dB) and the extra noise
$N_{e}=21.1$. All the solid curves are theoretical fitting according to different experimental parameters. \textbf{b}, The noise variance of the conditional
arbitrary quadrature components with the initial squeezing 3.9 dB
(antisqueezing 5.24 dB), the extra noise $N_{e}=1.75$ and the
optimal factor $g_{opt}=0.74$. \textbf{c}, The noise variance of the
conditional arbitrary quadrature components with the initial
squeezing 5.9 dB (antisqueezing 11.6 dB), the extra noise
$N_{e}=21.1$ and the optimal factor $g_{opt}=0.95$. The blue (light gray) curves
in (b) and (c) are the noise variance of the arbitrary quadrature
components of the lower sideband field of the squeezing light.
RBW=30 kHz, VBW=30 Hz and sweep time=500 ms. \label{Fig2} }
%}
\end{figure}

Fig. 2 shows the noise variance of the conditional quadrature phase
amplitude as the function of the factor $g^{2}$. Here, the noise of
the conditional quadrature phase amplitude is normalized to the SQL,
which is determined only with the LO1 (g=0) and injecting vacuum
field (blocking the squeezed light and signal field). When injecting
the squeezed light and given the intensity of the LO2 (given the
factor of g), the conditional arbitrary quadrature components are
measured by scanning the relative phase of $\theta$. The conditional
quadrature phase amplitude as the function of the factor $g^{2}$
(Fig. 2(a)) can be obtained by finding the minimum and maximum noise
variance from the measured arbitrary quadrature components, which
are good agreement with the theoretical calculation. Fig. 2(a) gives
two different squeezing (antisqueezing) with 3.9 dB (5.24 dB) and
5.9 dB (11.6 dB) respectively. Here, the extra noise $N_{e}$ of the
antisqueezed component can be calculated from the values of
squeezing and antisqueezing (see Appendix). Fig. 2(b) and (c) the
noise variance of the conditional arbitrary quadrature components
with the optimal factor $g_{opt}$ for two different input squeezing
with 3.9 dB and 5.9 dB respectively. Thus, the minimum conditional
variance is obtained with -1.5 dB for the initial squeezing of 3.9
dB and -3.1 dB for that of 5.9 dB. Here, the noise variances of the
arbitrary quadrature components of the lower sideband field of the
squeezing light (g=0) are constant and larger than SQL as the
function of the relative phase of $\theta$, which demonstrates that
one beam of an EPR entangled pair is thermal state.

\begin{figure}
\centerline{
\includegraphics[width=8cm]{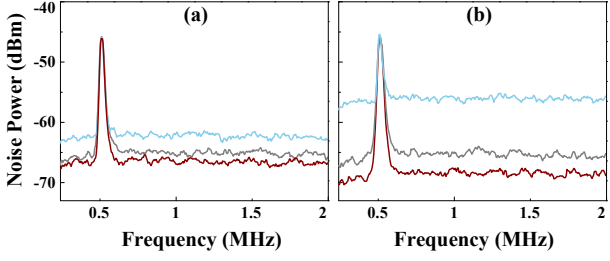}} \vspace{0.1in}
%\setlength{\columnwidth}{3.2in}
%\centerline{
\caption{\textbf{The noise spectra of the low-frequency signal with
500 kHz around the frequency $\omega_{0}$ by using broad squeezed
light and BLO.} \textbf{a}, The initial squeezing is 3.9 dB
($e^{-2r}=0.4$) and the extra noise is $N_{e}=1.75$. \textbf{b}, The
initial squeezing is 5.9 dB ($e^{-2r}=0.26$) and the extra noise is
$N_{e}=21.1$. The black (gray) curves are the noise spectrum of the signal
field with the input vacuum field and g=0. The blue (light gray) curves are the
noise spectrum of the signal field with the input squeezed field and
g=0. The red (dark gray) curves are the noise spectrum of the signal field with
the input squeezed field and $g_{opt}$. RBW=10 kHz, VBW=30 Hz.
\label{Fig3} }
%}
\end{figure}

Fig. 3 shows the enhanced sensitivity of low-frequency signal with
500 kHz around the frequency $\omega_{0}$ by using broad squeezed
light and BLO. When the vacuum field is injected, the noise floor
(black (gray) curve in Fig. 3) of the signal corresponds to SQL with g=0.
If the squeezed field is injected and g=0, the very noisy floor
(blue (light gray) curve in Fig. 3) is the noise variance of one beam of an EPR
entangled pair. When we choose the optimal factor $g_{opt}$, the
enhanced sensitivity of low-frequency signal is obtained and the
signal-noise ratio is improved with 1.5 dB for the initial broad
squeezing of 3.9 dB and 3.1 dB for that of 5.9 dB. Quantum advantage
resulting from the use of squeezed light is evaluated by comparing
signal-to-noise ratios in this work. Because of additional degrees
of freedom such as the optical gain g, a fair comparison between
quantum and classical light can be difficult. However, the quantum
noise floors of both the classical and the quantum approaches are
compared after independent factor over g in this works, because
changing g here does not change the signal level, as evidenced in
Fig. 3. In this scheme, one can avoid the low frequency technical
noise of the squeezed-light source by placing the signal and the
detection in one of the squeezed sidebands. This is useful only if
it is effectively the dominant source of noise. On the other hand,
the low frequency noise introduced by homodyne detection can not
avoid in the scheme. For example, amplitude noise will be rejected
by the balanced detector, up to the common-mode rejection power.
However, phase noise on will still creep in, particularly if the
squeezing is strong.

In conclusion, we have demonstrated a scheme of using a broadband
high-frequency squeezed vacuum to detect low-frequency signal beyond
the standard quantum limit. By means of the EPR entanglement of
upper and lower sideband of the broadband squeezed light, the
conditional variance in the low frequency band can be obtained by
BLO detection with subtracting the photocurrent of the upper
sideband beam from that of the lower sideband beam. Thus this scheme
does not need directly generate the squeezing in the low frequency
band. In addition, the BLO detection directly measures the signal
mapped on the sideband of the squeezed state, in this sense this
scheme can to some extent avoid the DC technical noise in the
traditional homodyne detection stemming from the light sources.

\begin{acknowledgments}
The authors would like to thank Yiqiu Ma for helpful discussion.
This research is supported by the MOST (Grant No. 2016YFA0301602),
NSFC (Grant No. 11234008, 11361161002,
11474188,11654002,6157127), Natural Science Foundation of Shanxi
Province (Grant No. 2015011007), Research Project Supported by
Shanxi Scholarship Council of China (Grant No. 2015-002).
\end{acknowledgments}
$^{\dagger}$Corresponding author email: jzhang74@sxu.edu.cn,
jzhang74@yahoo.com.

\begin{appendices}

\textbf {APPENDIX:The conditional squeezing with he extra noise of
the antisqueezed component }

Usually, the broadband squeezed light generated by OPO is not the
minimum uncertainty state, in which the antisqueezed component has
the extra noise. So the phase amplitudes of the output lower and
upper sideband fields of OPO with the extra noise can be given by

\begin{eqnarray}
\hat{X}_{-}^{s}(\Omega)&=&\hat{X}_{-}^{0}(\Omega)\cosh{r}+\hat{X}_{+}^{0}(\Omega)\sinh{r}+\frac{N_{X}}{2},\nonumber\\
\hat{Y}_{-}^{s}(\Omega)&=&\hat{Y}_{-}^{0}(\Omega)\cosh{r}-\hat{Y}_{+}^{0}(\Omega)\sinh{r}+\frac{N_{Y}}{2},\nonumber\\
\hat{X}_{+}^{s}(\Omega)&=&\hat{X}_{+}^{0}(\Omega)\cosh{r}+\hat{X}_{-}^{0}(\Omega)\sinh{r}+\frac{N_{X}}{2},\nonumber\\
\hat{Y}_{+}^{s}(\Omega)&=&\hat{Y}_{+}^{0}(\Omega)\cosh{r}-\hat{Y}_{-}^{0}(\Omega)\sinh{r}-\frac{N_{Y}}{2},
\end{eqnarray}
here,
$\langle\delta^{2}N_{X}\rangle=\langle\delta^{2}N_{Y}\rangle=N_{e}$.
The difference and sum of amplitude phase quadratures of the output
lower and upper sideband fields of OPO are obtained

\begin{eqnarray}
\hat{X}_{-}^{s}(\Omega)-\hat{X}_{+}^{s}(\Omega)&=&(\hat{X}_{-}^{0}(\Omega)-\hat{X}_{+}^{0}(\Omega))e^{-r},\nonumber\\
\hat{Y}_{-}^{s}(\Omega)+\hat{Y}_{+}^{s}(\Omega)&=&(\hat{Y}_{-}^{0}(\Omega)+\hat{Y}_{+}^{0}(\Omega))e^{-r},\nonumber\\
\hat{X}_{-}^{s}(\Omega)+\hat{X}_{+}^{s}(\Omega)&=&(\hat{X}_{-}^{0}(\Omega)+\hat{X}_{+}^{0}(\Omega))e^{+r}+N_{X},\nonumber\\
\hat{Y}_{-}^{s}(\Omega)-\hat{Y}_{+}^{s}(\Omega)&=&(\hat{Y}_{-}^{0}(\Omega)-\hat{Y}_{+}^{0}(\Omega))e^{+r}+N_{Y}.
\end{eqnarray}
The variances of the output lower and upper sideband fields of OPO
are expressed by
\begin{eqnarray}
\langle\delta^{2}X_{-}^{s}(\Omega)\rangle&=&\langle\delta^{2}X_{+}^{s}(\Omega)\rangle=\langle\delta^{2}Y_{-}^{s}(\Omega)\rangle=\langle\delta^{2}Y_{+}^{s}(\Omega)\rangle \nonumber\\
&=&\frac{e^{-2r}+e^{2r}}{2}+\frac{N_{e}}{4},
\end{eqnarray}
and the correlated variances are given by
\begin{eqnarray}
\langle\delta^{2}(\hat{X}_{-}^{s}(\Omega)-\hat{X}_{+}^{s})(\Omega)\rangle&=&\langle\delta^{2}(\hat{Y}_{-}^{s}(\Omega)+\hat{Y}_{+}^{s}(\Omega))\rangle=2e^{-2r},\nonumber\\
\langle\delta^{2}(\hat{X}_{-}^{s}(\Omega)+\hat{X}_{+}^{s})(\Omega)\rangle&=&\langle\delta^{2}(\hat{Y}_{-}^{s}(\Omega)-\hat{Y}_{+}^{s}(\Omega))\rangle=2e^{+2r}+N_{e}.
\end{eqnarray}

The variance of the conditional quadrature phase
$\hat{Y}_{B}(\Omega)$ and amplitude $\hat{X}_{B}(\Omega)$ detected
by the BLO with factor are expressed by
\begin{eqnarray}
\langle\delta^{2}\hat{Y}_{B}(\Omega)\rangle&=&\langle\delta^{2}(\hat{Y}_{-}^{s}+g\hat{Y}_{+}^{s})\rangle\nonumber\\
&=&\frac{1}{2}(e^{2r}+\frac{N_{e}}{2})(1-g)^{2}+\frac{e^{-2r}}{2}(1+g)^{2},\nonumber\\
\langle\delta^{2}\hat{X}_{B}(\Omega)\rangle&=&\langle\delta^{2}(\hat{X}_{-}^{s}+g\hat{X}_{+}^{s})\rangle\nonumber\\
&=&\frac{1}{2}(e^{2r}+\frac{N_{e}}{2})(1+g)^{2}+\frac{e^{-2r}}{2}(1-g)^{2}.
\end{eqnarray}
When we choose the optimized value of
$g_{opt}=(e^{2r}+N_{e}/2-e^{-2r})/(e^{2r}+N_{e}/2+e^{-2r})$, the
minimum conditional variance of the output lower sideband field is
obtained
\begin{eqnarray}
\langle\delta^{2}\hat{Y}_{B}^{opt}(\Omega)\rangle&=&\frac{2e^{-2r}(e^{2r}+N_{e}/2)}{e^{2r}+N_{e}/2+e^{-2r}}.
\end{eqnarray}
At the same time, we may give the quadrature amplitude
$\hat{Y}_{B}(\Omega)$ with the same condition is expressed by
\begin{eqnarray}
\langle\delta^{2}\hat{X}_{B}(\Omega)\rangle&=&\langle\delta^{2}(\hat{Y}_{-}^{s}+g\hat{Y}_{+}^{s})\rangle\nonumber\\
&=&\frac{1}{2}(e^{2r}+\frac{N_{e}}{2})(1-g)^{2}+\frac{e^{-2r}}{2}(1+g)^{2}.
\end{eqnarray}

\end{appendices}

\end{document}